\documentclass[aps,prb,floatfix,twocolumn,showpacs]{revtex4}

\usepackage{graphicx}
\usepackage{graphics}
\usepackage{dcolumn}
\usepackage{bm}
\usepackage[mathlines]{lineno}
\usepackage{amsmath}
\usepackage{amssymb}
\usepackage{scrextend}
\usepackage{hyperref}

\usepackage{color}%

\begin{document}
\title{Atomistic potential for graphene and other sp$^2$ carbon systems}

\author{Zacharias G. Fthenakis}
\affiliation{Institute of Electronic Structure and Laser, FORTH, Heraklion, Greece}
\affiliation{Department of Physics, University of South Florida, Tampa, Florida 33620, USA}
\author{George~Kalosakas}
\affiliation{Materials Science Department, University of Patras, Rio GR-26504, Greece}
\affiliation{Crete Center for Quantum Complexity and Nanotechnology (CCQCN), Physics Department, University of Crete GR-71003 Heraklion, Greece}
\author{Georgios D. Chatzidakis} 
\affiliation{Department of Physics, National Technical University of Athens, GR-15780 Athens, Greece}
\author{Costas~Galiotis}
\affiliation{Chemical Engineering Department, University of Patras, GR-26504, Rio, Greece}
\affiliation{ICE-HT/FORTH, PO Box 1414, GR-26504 Rio, Greece}
\author{Konstantinos~Papagelis}
\affiliation{Physics Department, University of Patras, GR-26504, Rio, Greece}
\affiliation{ICE-HT/FORTH, PO Box 1414, GR-26504 Rio, Greece}
\author{Nektarios N. Lathiotakis}
\affiliation{Theoretical and Physical Chemistry Institute, National Hellenic
Research Foundation, Vass. Constantinou 48, GR-11635 Athens, Greece  }

\date{\today}
\pacs{}

\begin{abstract}
We introduce a torsional force field for sp$^2$ carbon to augment an
in-plane atomistic potential of a previous work 
(Kalosakas et al, J. Appl. Phys. {\bf 113}, 134307 (2013)) so that it is 
applicable to out-of-plane deformations of graphene and related carbon materials. 
The introduced force field is fit to reproduce DFT calculation data of appropriately 
chosen structures. The aim is to create a force field that is as simple as possible
so it can be efficient for large scale atomistic simulations of various sp$^2$ carbon structures
without significant loss of accuracy. We show that the complete proposed potential reproduces
characteristic properties of fullerenes and carbon nanotubes. 
In addition, it reproduces very accurately the out-of-plane ZA and ZO modes of graphene's 
phonon dispersion as well as all phonons with frequencies up to 1000~cm$^{-1}$.
\end{abstract}
\maketitle

\section{Introduction}

Carbon nanostructures with predominant sp$^2$ bonds, like carbon fullerenes, nanotubes (CNTs) and graphene,
are in the center of scientific and technological interest for more than three decades~\cite{Kroto,Iijima,Novoselov666}.
This interest has increased significantly after the synthesis and identification of single layer  graphene
which has triggered an unprecedented focus of research on the material itself, its potential applications
and other two-dimensional materials~\cite{CastroNetoRMP,FerrariSSC,C4NR01600A,RuoffAM,GeimGrigorNat,2drevACSnano}.
The accurate, yet efficient, modeling of sp$^2$ bonded carbon at an atomistic level 
remains a great challenge for theory. Although there exist several atomistic 
models~\cite{Tersoff,Tersoff1,T2010,brenner,REAXFF,reaxff_gr,LCBOP,LCBOPII,AIREBO,Wei} 
and many of them have been proven
accurate in describing several properties of carbon based nanomaterials at a microscopic level,
there is a continuous need for as simple as possible and at the same time as accurate as possible models that 
could allow the simulation at a large, and increasing, scale~\cite{FasolinoNatMat,FasolinoPRL,Aluru,Buehler,peeters,arisJPCM,fthen_tomanek,ZhangNatCom,PCCP,yoon,aris2Dmat,meng,jain}. 

In a previous work \cite{kalosakas}, a force field for graphene 
was presented with terms depending only on atomic displacements within the graphene plane.  
In the present work, 
we extend that potential with the inclusion of torsional terms. The improvement 
achieved with these new terms is that it can be used to describe accurately 
out-of-plane distortions. For validation, we apply the present potential to
several characteristic test cases like the relative stability of fullerene isomers, 
the strain energy and the Young's modulus of CNTs and the phonon dispersion of graphene.   

The potential energy in atomistic simulations can be approximately expressed as the sum of several 
terms corresponding to specific geometric deformations originating from covalent,  electrostatic
or weak interactions. In the present, for $sp^2$ carbon systems, we consider only terms arising from covalent bonding.
Electrostatic terms are excluded since there is no charge localization on atoms, while for a single 
layer of sp$^2$ carbon atoms weak interactions can be neglected.
Therefore, we assume a deformation energy $U$ of the form
\begin{equation}
\label{eq:E}
 U=U_{\rm str} + U_{\rm bend} + U_{\rm tors}
\end{equation}
where $U_{\rm str}$ is a bond-stretching term, $U_{\rm bend}$ an angle-bending term and 
$U_{\rm tors}$ a term depending on out-of-plane torsional angle deformations. 
In a previous work~\cite{kalosakas},  parametric forms for the first two terms were presented,
derived through fitting to ab-initio calculations. The individual (for a single bond or angle)
bond-stretching and angle-bending terms, i.e. the contributions to $U_{\rm str}$ and
$U_{\rm bend}$ corresponding to a single bond or angle distortion, 
have respectively the forms~\cite{kalosakas}
\begin{equation}  \label{bsp}
V_s(r) = D \left[ e^{-a(r-r_0)}-1 \right]^2\,,
\end{equation}
with parameters $D=5.7 \; eV$, $a=1.96$~\AA$^{-1}$, and $r_0=1.42$~\AA, and
\begin{equation}  \label{abp}
V_b(\theta) = \frac{k}{2} \left( \theta-\frac{2\pi}{3} \right)^2 - \frac{k^\prime}{3} \left( \theta-\frac{2\pi}{3} \right)^3\,,
\end{equation}
with $k=7.0 \; {\rm eV}/{\rm rad}^2$ and $k^\prime=4 \; {\rm eV}/{\rm rad}^3$. Then the deformation energy terms of Eq.~(\ref{eq:E}) 
are $U_{\rm str} = \sum_i V_s(r_i)$ and $U_{\rm bend}=\sum_j V_b(\theta_j)$,
with $i$ and $j$ enumerating all the different bonds and angles, respectively.

This potential was proven useful in several cases, for example 
in reproducing accurately several elastic properties of graphene~\cite{kalosakas}.
However, it is only applicable to cases for which no out-of-plane deformations occur. 
For out-of-plane distortions, the bond-stretching and angle-bending terms alone are 
not sufficient to yield accurate results and augmentation of the force-field with
torsional terms is required (see for instance the discussion for C$_{40}$ isomers in 
Sec.~\ref{sec:fullerenes}). This is the main task of this work. Following the same recipe 
as in Ref.~\onlinecite{kalosakas} we assume a simple form for the individual torsional term
with parameters that are fitted to reproduce deformation energies from ab-initio calculations.
Our aim is to keep the potential as simple as possible, so it can be used in large scale molecular dynamics
or Monte Carlo atomistic simulations in graphene and other sp$^2$ carbon
materials, being at the same time as accurate as possible.

In order to check the efficiency of the proposed potential, we implemented it in 
LAMMPS computer code\cite{lammps_paper,lammps} 
and tested its speed compared with two popular atomistic models, namely 
Tersoff-2010\cite{T2010}, and LCBOP~\cite{LCBOP}. 
We simulated graphene with periodic boundary conditions assuming a cell of 1152 atoms. 
We found that our scheme is 3 and 4 times faster in this simulation than  Tersoff-2010 and  
LCBOP, respectively. Thus, the present potential will be a very good choice for extended sp$^2$ carbon systems. In case local phenomena are under study, for instance functionalization of such systems, then one expects that locally the structures depart from  sp$^2$ hybridization. In that case, the potential can be either modified locally to account more accurately for such effects, or replaced locally by a more accurate, but likely less effective scheme. 

This paper is organized as follows: In Section~\ref{tors_pot}, we describe the additional torsional terms
and the procedure of fitting, i.e. optimizing the associated parameters. Full details for
this procedure are given in a separate report~\cite{torsional_p}. Then, in Section~\ref{pot_valid}, we 
demonstrate the efficiency and the accuracy of this potential in specific applications:
energetics of fullerenes (Sec.~\ref{sec:fullerenes}),
the strain energy and the Young's modulus of CNTs (Sec.~\ref{sec:nanotubes}),
and the phonon dispersion of graphene (Sec.~\ref{sec:phonons}).
Conclusions are included in Sec.~\ref{sec:conclusion}.

\section{Torsional Force Field\label{tors_pot}}

\begin{figure}[!tb]
\begin{center}
\begin{tabular}{c}
\includegraphics[width=0.45\columnwidth,clip]{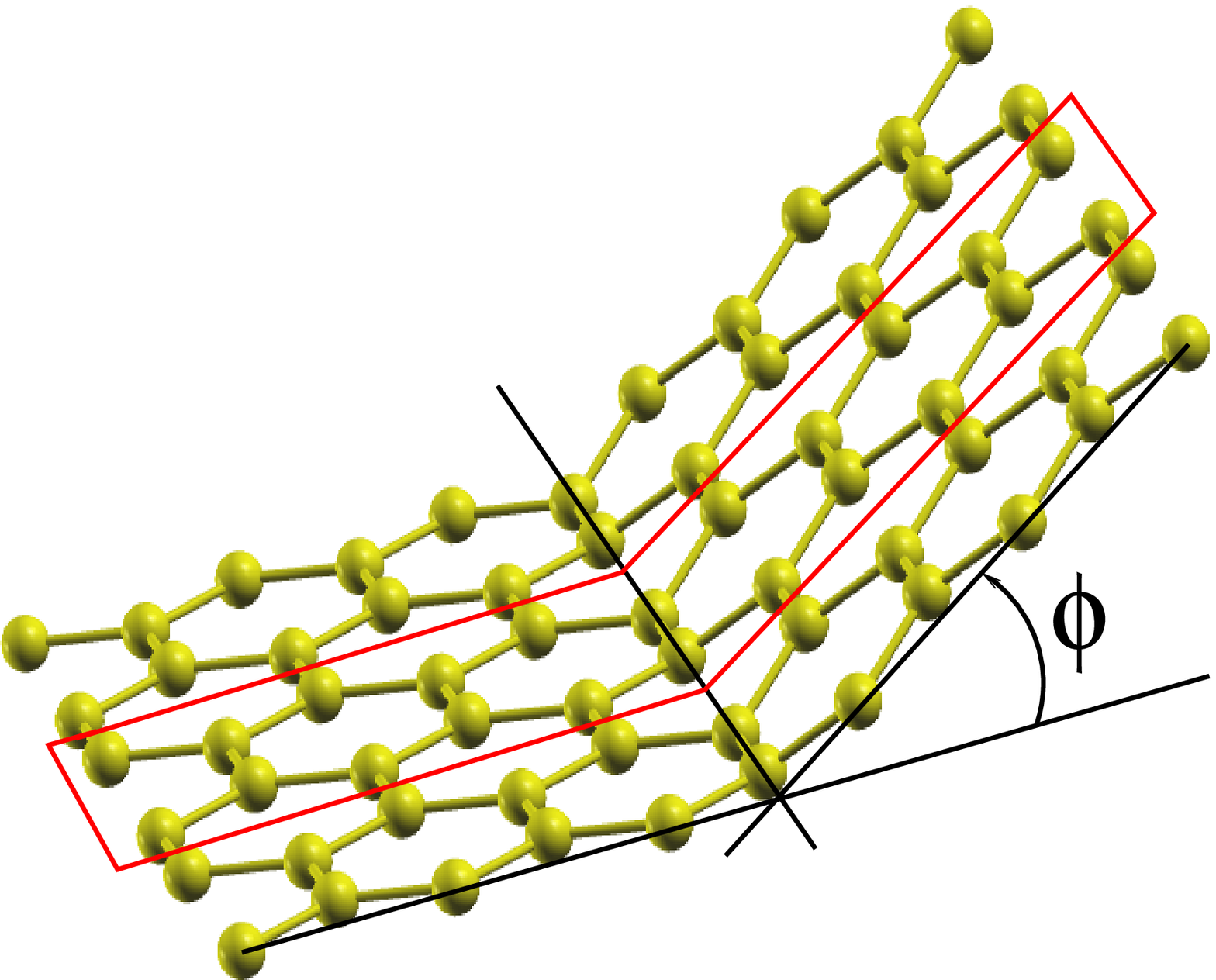} \\
\includegraphics[width=0.50\columnwidth,clip]{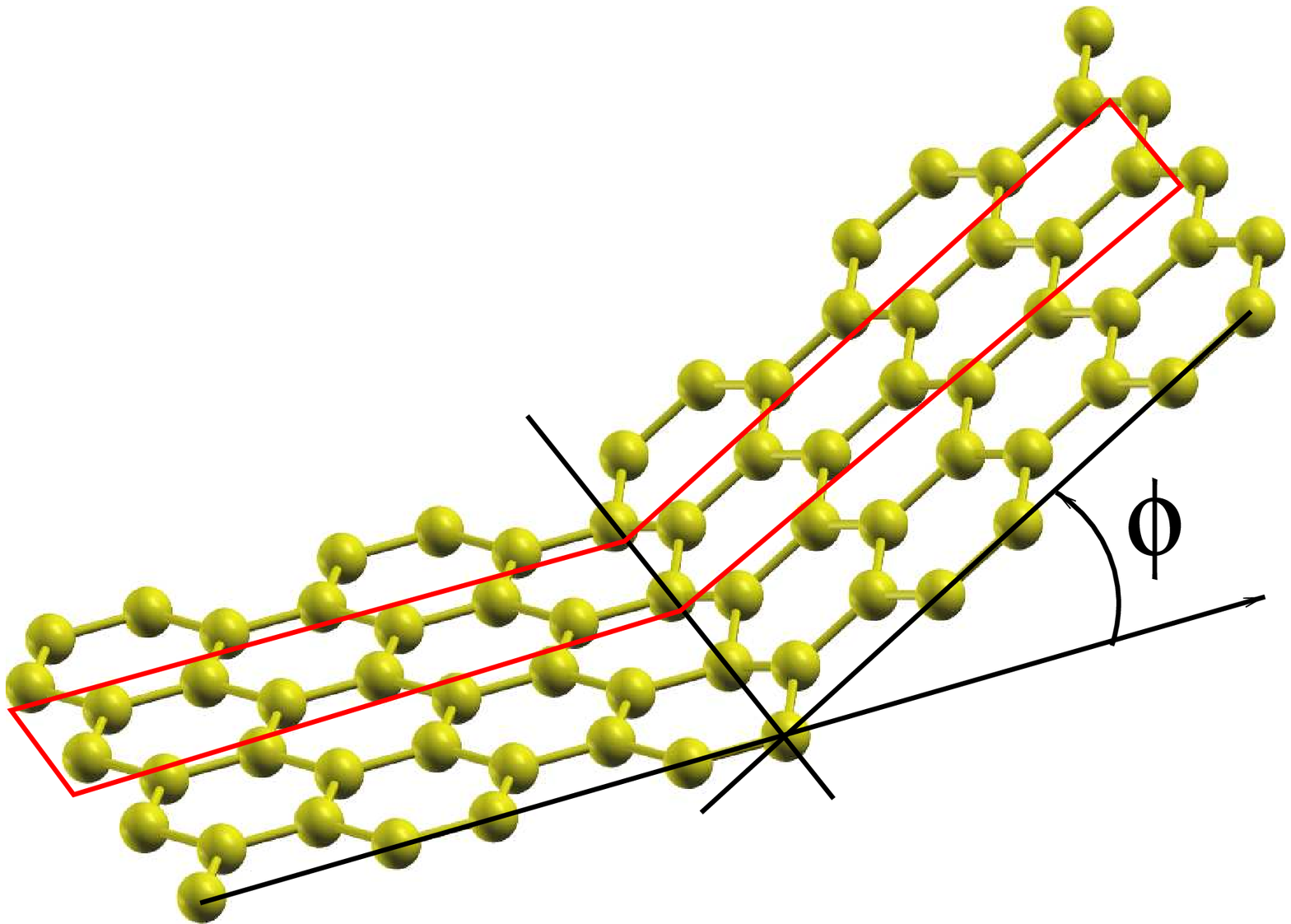} \\
\end{tabular}
\end{center}
\caption{ The structures simulated for folding around an axis along the armchair 
(top) and along the zig-zag direction (bottom). Unit cells are shown in red frame.  }
\label{fig:struc_E}
\end{figure}

In order to fit the parameters of a torsional term, we calculated 
ab-initio the total energies of two structures of a single graphene layer folded by an angle $\phi$
around an axis lying along either an armchair or a zig-zag direction. These structures are shown in 
Fig.~\ref{fig:struc_E} and the corresponding unit cells are also included. The structures are 
periodic along the direction of the folding axis but finite along the vertical direction. 
In other words, they are ribbons that fold around their middle line.
Due to periodicity, a minimal size for the unit cell was adopted in the folding 
direction, and a large one in the vertical in order to avoid edge effects as much as possible.
For the armchair-axis folding (Fig.~\ref{fig:struc_E} top), the adopted 
unit cell contains 22 atoms, while for the zig-zag one (Fig.~\ref{fig:struc_E} bottom)
17 atoms. In that way, in the case of armchair folding axis, for all atoms located on this axis,
 all neighbors up to the 5th in the vertical zig-zag direction are included in the simulation
 (see Fig.~\ref{fig:struc_E}~top). In the case of the zig-zag folding axis, with a thiner unit 
cell along the edge, all neighbors up to the 8th in the vertical armchair direction are 
included (Fig.~\ref{fig:struc_E}~bottom). Finally, the unit cells also contain sufficient 
empty space in both vertical directions outside the ribbons.
Calculations were performed at the level of generalized gradient approximation, 
Perdew-Burke-Ernzernhof functional\cite{PBE}, of density functional theory (DFT) using the
Quantum-Espresso periodic code\cite{QE-2009}.   
We used the same pseudopotential\cite{pseudo} as in Ref.~\onlinecite{kalosakas} and plane-wave
cutoffs 40 and 400~Ry, for the wave-function and density, respectively.

Obviously, the deformations shown in Fig.~\ref{fig:struc_E} are rather complex. They 
involve several individual torsional terms of different torsional angles and, in addition, 
angle-bending terms. 
This complicates the fitting process which is described briefly here.
In order to fit an analytic form for the individual torsional term, 
it is necessary to separate the total torsional energy, i.e. to 
remove all the angle-bending contributions from the total deformation energy.
This was achieved by identifying and expressing analytically, in terms
of $\phi$, all the changes in bending angles 
induced by the folding. Then the angle-bending expressions of Eq.~(\ref{abp}) were used to
account for the individual angle-bending terms which were subsequently subtracted
from the points of the total deformation energy to obtain the remaining ``pure" torsional energy.
More details are presented elsewhere~\cite{torsional_p}.
 
In order to proceed with the fit, we express the total torsional energy 
analytically in terms of $\phi$ including all individual torsional terms that
contribute. Thus, we have identified all torsional angles altered by the
folding and express them in terms of $\phi$. Finally we have to assume a
fitting form for the individual torsional terms, i.e. the 
contribution to $U_{\rm tors}$ corresponding to a single torsional angle $\omega$, 
 and we chose the following
that respects rotational symmetry\cite{jensen}:
\begin{eqnarray}
V_t(\omega) &=& \frac{1}{2} V_1 \left[ 1 + \cos ( \omega ) \right] \nonumber \\
 &+& \frac{1}{2} V_2 \left[ 1 - \cos (2 \omega ) \right] \label{eq:tors0} \,,
\end{eqnarray}    
where $V_1$, $V_2$ are parameters to be optimized. 

The fitting can be performed for several choices regarding the range of $\phi$, $[0, \phi_{\rm max}]$, 
and we have investigated three different ones~\cite{torsional_p}. 
The optimal values we found are $V_1 \sim 2\times10^{-4}$~eV and $V_2\approx0.23$~eV.
Due to the small value of $V_1$, we can neglect the
first term in Eq.~(\ref{eq:tors0}) simplifying further the potential expression. Thus, our final
proposition for the individual torsional term is
\begin{equation}
\label{eq:tors1}
V_t(\omega) = \frac{1}{2} V_2 \left[ 1 - \cos (2 \omega ) \right], \quad V_2=0.23\:eV\,.
\end{equation}
The fitted function reproduces reasonably well the DFT data up to folding angles of 30$^{\circ}$ 
($\approx$0.5~rad)\cite{torsional_p}.

\section{Potential Validation\label{pot_valid}}

The important extension provided here regarding the in-plane potential presented in Ref.~\onlinecite{kalosakas}
is the incorporation of the out-of-plane torsional term. This term is crucial as it permits calculations of out-of-plane
deformations in planar graphenes and also simulations of non-planar sp$^2$ carbon structures, like for instance
fullerenes and CNTs.
In this section, the full potential that includes the in-plane terms given by Eqs.~(\ref{bsp}) and (\ref{abp})
and the torsional term of Eq.~(\ref{eq:tors1}) is applied to several
test  cases to check whether various experimental or accurate ab initio theoretical results  are reproduced. 
We focus on examples for which the correct description of out-of-plane deformations is important.
Thus, we test our potential against (a)  the energy of fullerene isomers, (b) the energy and the Young's modulus
of nanotubes with different chirality $(n,m)$, and (c) the phonon dispersion relations of graphene 
focusing on the out-of-plane acoustic (ZA) and optical (ZO) modes.

\subsection{Fullerenes\label{sec:fullerenes}}

Apart from the well known icosahedral C$_{60}$ fullerene \cite{Kroto},
many other fullerenes exist. They may have different number of atoms and 
their pentagonal and hexagonal rings may be arranged differently \cite{atlas}. 
A C$_N$ fullerene is composed of 12 pentagonal and $N/2-10$ hexagonal
rings, where $N$ is an even number with $N\ge 20$. Due to the different 
arrangement of the pentagonal and hexagonal rings in a C$_{N}$ fullerene, 
many C$_N$ fullerene isomers exist, the number of which rise exponentially\cite{atlas} with $N$.

\begin{figure}[!tb]
\includegraphics[width=0.95\columnwidth,clip]{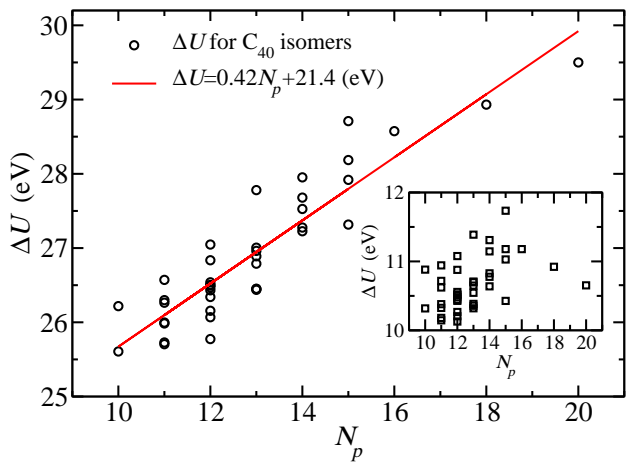}
\caption{
The energy of C$_{40}$ fullerene isomers, calculated with the present potential, with 
respect to the energy of 
graphene $\Delta U=U-U_{\rm graph}$ as a function of their pentagon adjacencies $N_p$.
The inset shows the corresponding results when the out-of-plane torsional term of the
potential is neglected.}
\label{fig:figure1}
\end{figure}

Albertazzi et al\cite{Fowler} showed that the energy $U$ of C$_{N}$ fullerene 
isomers rise almost linearly with the number $N_p$ of pentagon adjacencies.  
In their study they calculated the energy of the forty C$_{40}$ isomers
using 12 different methods, from molecular mechanics to 
very accurate ab-initio methods. The slopes of the linear relations depend on the
method used, and vary between 99.5~kJ/mol (for the ab-initio methods)
to 24.4~kJ/mol (for the molecular mechanics methods).
In the present study we repeat these calculations for the 
energy of the forty C$_{40}$ isomers, using our potential. 
In Fig.~\ref{fig:figure1}, we show the energy $\Delta U$ of these isomers 
with respect to the energy $U_{\rm graph}$ of graphene as a function of $N_p$,
where $U_{\rm graph}=40\:U_{\rm coh}^{\rm graph}$ and $U_{\rm coh}^{\rm graph}$ is the 
cohesive energy of graphene.
As one can see, the energy of the isomers rise linearly with the number $N_p$ of their
pentagon adjacencies in accordance with the results of Ref.~\onlinecite{Fowler}.
Moreover,  the energetically optimum isomer (isomer number 40:38, according to the isomer
enumeration provided by Ref.~\onlinecite{atlas}) is the same as the one found by Albertazzi et al.
using most of the 12 methods, including the ab-initio~\cite{Fowler}.
Using a least squares fitting we calculate the slope $a$ and the intercept  $b$ of the relation
$\Delta U=aN_p+b$. The values we found are  $a=0.42$~eV (or $a=40.5$~kJ/mol), 
and $b=21.4$~eV with standard error of estimate 0.53~eV.
The obtained value for $a$, is within the range of slopes found by  Albertazzi et al.

Using our potential we found that the energy of
the icosahedral C$_{60}$ fullerene with respect to the energy of graphene is 
$\Delta U(C_{60})=23.4$~eV or 
$\Delta U(C_{60})/N=0.39$~eV/atom, which is consistent with the corresponding experimentally 
obtained energy value $0.41 \pm 0.02$~eV/atom~\cite{flakes,Chen}, and the theoretically
obtained value 0.38~eV/atom using the DFT method at the GGA/PBE level \cite{fullerenes_2016}.
The corresponding energy found from the optimum C$_{40}$ isomer 
(isomer 40:38 of Ref.~\onlinecite{atlas}, with $N_p=10$)
is 0.64~eV/atom, in accordance with the value 14.6~kcal/mol (0.63~eV/atom),
that can be obtained from Ref.~\onlinecite{fullerenes_2016}. In that work, the energies of both
the optimum C$_{40}$ isomer and graphene are given with respect to that of the icosahedral C$_{60}$
obtained with the DFT/PBE method.

Neglecting the torsional terms of the Eq.~(\ref{eq:tors1}), we obtain
the energy $\Delta U$ of the C$_{40}$ fullerene isomers that is shown in the inset of 
Fig.~\ref{fig:figure1}, as a function of $N_p$.
As one can see, the linear relation between $\Delta U$ and $N_p$ is not reproduced
and the energy values $\Delta U$ range between 10 and 12~eV. This range is much smaller than
that obtained when torsional terms are included (between 25 and 30~eV).
Moreover, the energetically more favorable isomer found when torsional terms are
neglected is the 40:40 isomer, with $N_p=12$, and not the 40:38 obtained when terms are 
included. As for the icosahedral 
C$_{60}$, its energy without including the torsional term, is $\Delta U(C_{60})=9.9$~eV or 
$\Delta U(C_{60})/N=0.17$~eV/atom, which is much smaller than the experimental value of 0.41~eV.
This clearly shows the importance of the torsional terms of the proposed potential
for the accurate prediction of the energetics of the fullerene structures.

\begin{figure}[!tb]
\includegraphics[width=0.95\columnwidth,clip]{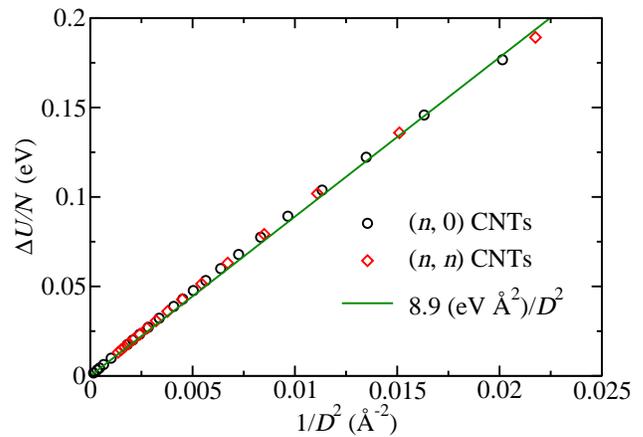}
\caption{The energy per atom $\Delta U/N$ with respect to the energy
of graphene, for the $(n,n)$ and $(n,0)$ CNTs (diamonds and circles, respectively) versus $1/D^2$, 
where $D$ is the nanotube diameter. The solid line represents a fitting with the formula $\Delta U/N=C/D^2$ (see text).}
\label{fig:figure2}
\end{figure}

\subsection{Carbon Nanotubes\label{sec:nanotubes}}

It has been shown that the strain energy per atom  $\Delta U/N=U/N-U^{graph}_{coh}$
of a  CNT, i.e. the energy per atom $U/N$ with respect 
to the cohesive energy of graphene $U^{graph}_{coh}$, depends on its diameter $D$. 
Tibbetts \cite{TIBBETTS}, using continuum elasticity
theory, showed that the strain energy per atom of a CNT is given by
$\Delta U/N = Ed_0^3S/(6D^2)$, where $E$ is the Young's modulus of the 
CNT, $S$ the area per atom and $d_0$ the interlayer separation in graphite.
Assuming that $E$ and $S$ are constants, the strain energy $\Delta U/N$ has 
a $D^{-2}$ dependence, i.e. $\Delta U/N=C/D^2$, where $C$ is a constant.

\begin{figure}[!tb]
\includegraphics[width=0.95\columnwidth,clip]{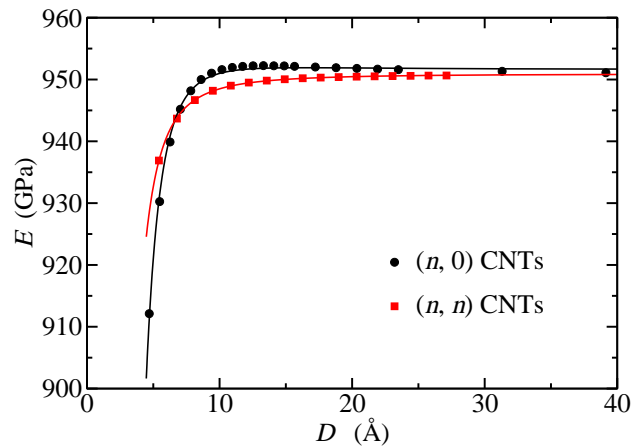}
\caption{Young's modulus, $E$, for the $(n,n)$ and $(n,0)$ CNTs (squares and circles, respectively)
as a function of their diameter $D$. Points show numerical results and solid lines fittings with
analytical relations (see text).}
\label{fig:figure3}
\end{figure}

The fitting value of $C$ varies, depending on the method used for the energy calculations.
Using the Tersoff \cite{Tersoff} potential,  Sawada and Hamada \cite{Sawada} found $C=5.64$~eV \AA$^2$,
while Tersoff\cite{Tersoff_CNTs} estimated $C=5.36$~eV \AA$^2$.
Zhong-can et al. \cite{Zhong-can} using continuum elasticity found $C=6.12$~eV \AA$^2$. 
Based on tight binding calculations, Xin et al. \cite{Xin} found $C=5.76$~eV \AA$^2$,
Molina et al. \cite{Molina} estimated $C=5.64$~eV \AA$^2$, while Hernandez et al. \cite{Rubio} 
found $C=8.7$~eV \AA$^2$ for the $(n,0)$ CNTs, and $C=8.1$~eV \AA$^2$ for the $(n,n)$ CNTs.
Using a tight binding density functional method, Adams et al. \cite{Adams} calculated
$C=8.46$ and 8.37~eV \AA$^2$ for the $(n,n)$ and the $(n,0)$ CNTs respectively.
S\'anchez-Portal et al. \cite{Portal}, performing DFT calculations in the LDA level\cite{siesta1,siesta2},
found $C=8.00$~eV \AA$^2$ for the $(n,n)$, while for the (8,4) and (10,0) CNTs they found
slightly larger values, $C=8.60$ and 8.64~eV~\AA$^2$, respectively. 

In Fig.~\ref{fig:figure2}, we show (with points) the energy $\Delta U/N$
as a function of $1/D^2$ of various optimized $(n,n)$ and $(n,0)$ CNTs,
for $D>6$~\AA, using the model potential presented here. 
As we can see, there is an almost linear relation between $\Delta U/N$
and $1/D^2$, in agreement with all the previous studies.
Fitting a linear function of $1/D^2$ to these energy values, we find that both $(n,0)$ and $(n,n)$ CNTs
can be approximately described by $\Delta U/N=C/D^2$, where $C=$8.9~eV~\AA$^2$.
This fitting is shown by continuous line in Fig.~\ref{fig:figure2}.
The obtained value of $C$, using the force field presented here, is close to the 
tight binding values of Hernandez et al \cite{Rubio}, the tight binding
density functional values of Adams et al \cite{Adams} and the DFT/LDA values of 
S\'anchez-Portal et al \cite{Portal}. This value is consistent to the
results obtained with the more accurate methods, in contrary to the corresponding results of other,
less accurate atomistic models.

Note a very small discrepancy between this analytical relation and the numerical data for intermediate values
of $D$ in Fig.~\ref{fig:figure2}, revealing that a $1/D^2$ dependence of the energy is not so precise. 
A more accurate relation for the energy per atom $\Delta U/N$ should include an additional
term depending on $1/D^4$, as reported by Kanamitsu and Saito \cite{Kanamitsu}.
Including this term in our fitting, we find
$\Delta U/N =9.68 (\mbox{eV \AA}^2) / D^2 - 45.3(\mbox{eV \AA}^4) / D^4 $.

Furthermore, we calculate the Young's modulus $E$ of the $(n,0)$ and $(n,n)$ CNTs
using the potential proposed here, through the relation
\begin{equation} 
E=\frac{1}{2} \frac{\partial^2 U}{\partial x^2} \frac{l}{\pi r d_0},
\end{equation}
where $U$ represents the deformation energy of a CNT segment with length $l$
deformed by $x$ ($x \equiv \delta l$), $r$ is the CNT radius and $d_0=3.34$~\AA\  is the width of the 
nanotube wall, where we have adopted  the convention that the width $d_0$
of the nanotube is the same as the interlayer separation  of graphite.
In Fig.~\ref{fig:figure3}, we plot the calculated $E$ values against the CNT diameter $D$.
As one can see, our model predicts a rapid increase of the Young's modulus $E$ of CNTs
as a function of their diameter $D$, for diameters up to 8~\AA. For larger diameters, 
as the strain due to rolling drops, $E$ remains almost constant with a value
equal to that of graphene. This result is 
in accordance with earlier studies using other model potentials 
\cite{Meo, Chang, Popov}, as well as tight binding \cite{Goze} and ab-initio  calculations \cite{Rubio}.
Moreover, we fit a quadratic function of $1/D^2$ to these data. The fitting functions for $(n,0)$ and $(n,n)$
CNTs are  $E=950+175 /D^2-23500/D^4$ and $E=950-175/D^2-7080/D^4$, respectively, where
$E$ is given in GPa and $D$ in \AA. These analytical relations are also shown in Fig.~\ref{fig:figure3}
with lines and we see that these functions perfectly fit to the calculated $E$ values.
It is worth noting that the corresponding Young's modulus
value predicted for graphene using our model potential is $E \approx 950$~GPa.

\subsection{Phonon dispersion relations of graphene\label{sec:phonons}}

\begin{figure}[!tb]
\includegraphics[width=0.95\columnwidth,clip]{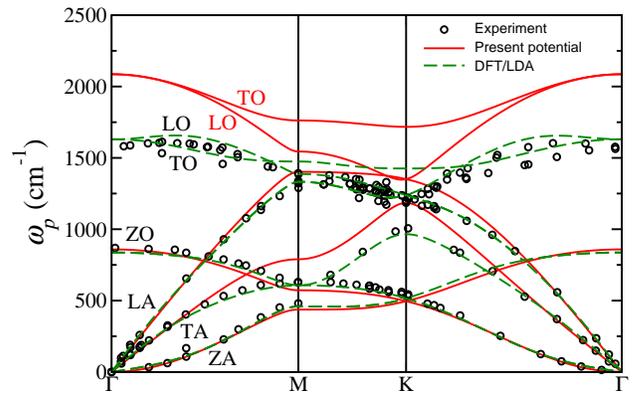}
\caption{Phonon dispersion relation of graphene along the 
$\Gamma$MK$\Gamma$ path calculated using the potential presented here
and the DFT method at the LDA level, compared with experimental data (points) of the phonon dispersion 
of graphite\cite{mohr,maultzch}.}
\label{fig:figure4}
\end{figure}

As another test of the force field presented here, we calculate 
the phonon dispersion relation of graphene along the $\Gamma$MK$\Gamma$ path
and compare it with experimental values~\cite{mohr,maultzch} as well as DFT calculations. 
The phonon dispersion relations were obtained using a hand-made computer code which
calculates the Hessian matrix
as the derivative of the atomic forces by inducing small perturbations of the atomic
positions.

DFT phonon calculations were performed at the LDA level employing the 
Perdue and Zunger\cite{PZ} exchange and correlation functional using
the SIESTA code \cite{siesta1, siesta2}.
We also used norm-conserving Trullier and Martins pseudopotenitals\cite{TM_pseudo}, 
and 10$\times$10$\times$1 Monkhorst Pack k-point grid. As for the basis set, we used a
double-$\zeta$ polarized basis set of atomic orbitals with a 100 Ry energy
cut-off.  For the phonon band structure 
calculations we used the ``vibra" utility of SIESTA.

\begin{figure}
\begin{center}
\includegraphics[width=0.85\columnwidth,clip]{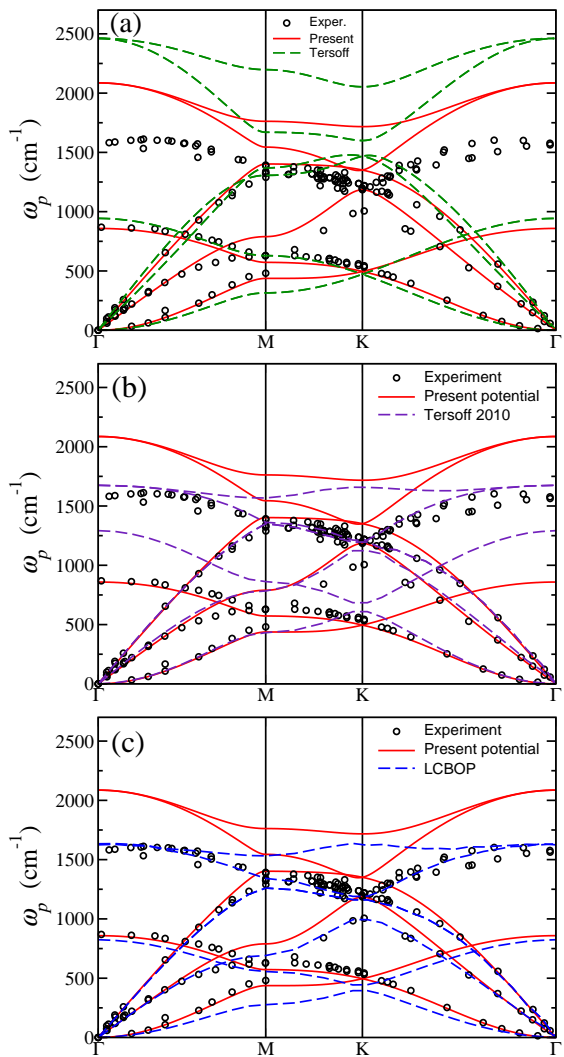}
\end{center}
\caption{Phonon dispersion relation of graphene along the 
$\Gamma$MK$\Gamma$ path calculated using the potential presented here compared with 
(a) the Tersoff, (b) the Tersoff-2010, and (c) the LCBOP potentials. The
experimental points of the phonon dispersion of 
graphite\cite{mohr,maultzch} are also included.}
\label{fig:comp_pot}
\end{figure}

The phonon dispersion calculated with our potential and DFT are shown in
Fig.~\ref{fig:figure4} together with the experimental data.
For comparison with known, widely used atomistic potentials, we show in Fig.~\ref{fig:comp_pot},
the dispersion obtained using our potential together with that using the Tersoff~\cite{Tersoff,Tersoff1},
a reparameterized Tersoff (Tersoff-2010)~\cite{T2010}, and LCBOP~\cite{LCBOP} potentials, in panels
(a)~-~(c), respectively. In order to quantify the performance of different schemes in phonon dispersion calculations,
we show in Table~\ref{tab:rmsd_phon} the RMSD error (root of the mean squared deviation with respect to
experimental values)  for all calculation schemes shown in Figs.~\ref{fig:figure4} 
and \ref{fig:comp_pot}. 
The numerical data for the dispersions obtained by the Tersoff, Tersoff-2010, and LCBOP potentials are taken from
Ref.~\onlinecite{koukaras}. As seen in Figs.~\ref{fig:figure4}, \ref{fig:comp_pot} and
Table~\ref{tab:rmsd_phon}, in general, DFT reproduces more accurately phonon dispersions than any
of the considered classical atomistic potentials, as expected, 
despite the fact that the simple LDA approximation was used. However, LO branch is an exception 
for which LCBOP and Tersoff-2010 seem to perform better.

\begin{table*}
\setlength{\tabcolsep}{8pt}
\caption{The RMSD error (in cm$^{-1}$) of the phonon dispersion of graphene for each phonon mode 
separately (first to sixth row) calculated with the present force field, Tersoff, Tersoff-2010, and LCBOP atomistic potentials,
as well as using DFT. The error 
is calculated with respect to the experimental points of the phonon dispersion of 
graphite\cite{mohr,maultzch}. We also include the RMSD error for
(i) the overall data of all modes (seventh row), (ii) all but excluding  LO and TO modes (eighth row), (iii) the out-of-plane ZA and
ZO modes (ninth row), and (iv) all phonons with low frequencies $\omega_p < 1000$~cm$^{-1}$ (tenth row).
}
\begin{center}
\begin{tabular*}{\textwidth}{@{\extracolsep{\fill}}lccccc}
\hline
   & Present  potential& Tersoff ($^{\alpha}$)
& Tersoff-2010 ($^{\alpha}$)
& LCBOP ($^{\alpha}$)
& DFT/LDA \\
\hline
ZA & \ 25.7    &  \ 71.5  &   \  38.8      & \ 99.9   &  \ 18.6   \\
ZO & \ 49.4    &  \ 37.5  &    238.8     & \ 97.8    &  \ 26.6   \\
LA & \ 97.1    &  137.6 &    \  21.3     & \ 32.4    &  \ 21.2   \\
LO & 273.7   &   773.1 &    \  36.1     & \ 32.8    & 130.4   \\
TA & \ 86.2    &   325.5 &   \   80.2     & \ 26.0    &  \ 18.0   \\
TO & 446.0   &   479.7 &   294.7      & 264.3   & \ 46.7    \\ \hline
{\bf All modes}& 235.7   &   472.2 &   149.8       & 118.1   & \ 72.9    \\ \hline
{\bf All modes, excluding LO and TO} & \ 75.3 & 192.4 & 121.2 & \ 67.1 & \ 21.2 \\ \hline
{\bf Out-of-plane modes: ZA and ZO} & \ 40.1 & \ 56.0 & 176.4 & \ 98.8 & \ 23.2 \\ \hline
{\bf Small frequencies ($< \mbox{10}^{3}\mbox{  cm}^{-1}$)} & \ 55.5 & 191.4 & 132.8 & \ 71.9 & \ 19.5 \\
\hline
\multicolumn{6}{l}{($^{\alpha}$) RMSD values are taken from Ref.~\onlinecite{koukaras}. } 
\end{tabular*}
\end{center}
\label{tab:rmsd_phon}
\end{table*}

Clearly the present potential, despite its simplicity,
is quite accurate in reproducing the phonon dispersion for all modes in the regime 
of small frequencies. Its performance in the linear regime 
of the LA and TA acoustic modes is excellent and in very close agreement with the experimental and 
the DFT results. This is demonstrated in Table~\ref{tab:rmsd_phon} where we show the RMSD errors
only for those phonons that their experimental frequency is smaller than 1000 cm$^{-1}$ 
(see last row).
As we see, in this regime, only the DFT results are superior to the present potential, while
those of the other atomistic potentials, including the two recent ones, are less accurate.
The only exception from the good performance in this  regime is the upper part of the TA mode,
where the results of the present potential seem to deviate from the experimental data.

Regarding the ZO and ZA modes (the lower optical and acoustic modes, respectively),
representing the out-of-plane phonons in graphene, the performance of the present potential
is quite satisfactory, and substantially better than any of the other atomistic potentials we
discuss here. This is quantified in
Table~\ref{tab:rmsd_phon}, where we include the RMSD errors for these modes (see ninth row).
Thus, in conclusion, the out-of-plane terms introduced in the present work, augmenting the
in-plane terms introduced in Ref.~\onlinecite{kalosakas}, add to the potential the capability
to calculate very accurately the out-of-plane phonon modes ZA and ZO of graphene.

Overall, the potential presented here is superior than the
Tersoff potential. This is the case for almost all modes as shown in Table~\ref{tab:rmsd_phon}
except ZO, for which, although our potential is very accurate, Tersoff potential performs
surprisingly well. Compared to the two more recent potentials, i.e. Tersoff-2010 and LCBOP,
the present potential shows a worse overall agreement with experiment. 
However, the overall performance of the present potential is negatively affected mainly by its 
inability to reproduce the LO and TO modes and, to a lesser extend, the high frequency regimes of TA mode.
These phonon frequencies are substantially overestimated. Its failure,
however, is less dramatic than Tersoff's potential. 
We mention that these modes are rather unaffected by
the out-of-plain torsional terms introduced in the present work. Instead, they depend strongly
on the in-plane bond-stretching and angle-bending terms. A possible future improvement would be
to extend the present model by including more distant stretching interactions (than just first neighbors)
or introducing mixed  stretching-bending terms. Such terms can also be fitted to ab-initio data 
similarly to the terms already included in the present potential. 

It should be stressed that graphene possesses anomalous optical-phonon dispersion at the 
$\Gamma$ and K points since alterations of the electronic screening occurs  for the 
atomic vibration at these particular points\cite{PhysRevLett.93.185503,FerrariSSC}
There exist Kohn anomalies at these points
reducing the frequency of the LO branch by tens or hundreds cm$^{-1}$. Our model does not
take into account such effects.

Note that there exist graphene properties, like the lattice thermal conductivity at room temperature,
that seem to entirely determined by the lower frequency and the out-of-plane phonon modes \cite{tcJAP14,tcJAP15}.
In particular, it has been found that even for higher temperatures, up to 800-1000~$K$,
the contribution of the high frequency LO and TO modes in the thermal conductivity does not exceed
a value of around 5\% \cite{tcJAP14,tcJAP15}. Our proposed potential may combine improved efficiency
and accuracy for studying such properties of graphene using atomistic simulations.

\section{\label{sec:conclusion} Conclusions}

In conclusion, we introduced a torsional term for sp$^2$ carbon systems in order to complement
the in-plane force field which was presented previously~\cite{kalosakas}. For this task, we 
performed DFT calculations for two different, appropriately chosen, graphene-nanoribbon structures 
that were folded around their middle line. The torsional terms were fitted to reproduce the 
deformation energy of these structures as a function of the folding angle. 
The proposed torsional potential has the simple form of Eq.~(\ref{eq:tors1}).Our aim was to keep the
proposed force field as simple as possible, targeting computational efficiency, for large scale
simulations. Indeed, in a test simulation with LAMMPS, our force field was found 
to be 3 and 4 times faster than Tersoff/Tersoff-2010 and LCBOP potentials, respectively.

The full proposed potential was tested in several characteristic cases. More specifically, we demonstrated that, with the
inclusion of torsional terms, the linear dependence of the energy of C$_{40}$ fullerene isomers on the
number of adjacent pentagons is obtained, while the energy of the more favorable C$_{60}$
and C$_{40}$ fullerenes is accurately reproduced.
Then, we showed that, in the case of carbon nanotubes, the proposed potential reproduces the 
$1/D^2$ dependence of the strain energy on the nanotube diameter $D$, while its predictions for the
Young's moduli of nanotubes as a function of their diameter are in accordance with existing calculations. 

Finally, we calculated the phonon dispersion of graphene and compared with other atomistic force fields and
DFT, as well as to experimental data.
The performance of the present potential for the phonon modes can be separated in two 
frequency regions. For the high frequency phonons ($\omega_p> 1000$~cm$^{-1}$), especially for LO, 
TO modes, our potential overestimates phonon frequencies substantially, however, it is still better than Tersoff potential.
In the low frequency regime, $\omega_p< 1000$~cm$^{-1}$ (including the
behavior of acoustic modes close to the $\Gamma$ point),
its accuracy is quite satisfactory with results closer to the experimental data and the DFT values
as compared to other widely used atomistic potentials. We stress the 
success of the present potential in reproducing the ZA and ZO out-of-plane phonon modes of 
graphene very accurately, and in better agreement with experiment compared to other recent
and popular classical force fields. This close agreement is a result of
the out-of-plane torsional terms introduced in the present work. 

{\it Acknowledgements:}
We acknowledge helpful discussions with E. N. Koukaras and A. P. Sgouros.
We also thank A. P. Sgouros for performing the efficiency comparison of the presented potential in LAMMPS.
The research leading to the present results has received funding from Thales project
``GRAPHENECOMP'', co-financed by the European Union (ESF) and the Greek Ministry of Education 
(through ESPA program). NNL  acknowledges support from the Hellenic Ministry of Education/GSRT (ESPA), through 
``Advanced Materials and Devices'' program (MIS:5002409) and EU H2020 ETN project ``Enabling Excellence'' Grant 
Agreement 642742. The FORTH/ ICE-HT contributors acknowledges the support of the European Research Council 
Advanced Grant ``Tailor Graphene'' (no. 321124, 2013-2018).


\end{document}